\documentclass[3p,times,procedia]{elsarticle}
\usepackage{ecrc}

\volume{00}

\firstpage{1}

\journalname{Nuclear Physics A}

\runauth{C. Chattopadhyay et al.}

\jid{nupha}

\jnltitlelogo{Nuclear Physics A}

\usepackage{amssymb}

\usepackage[figuresright]{rotating}

\usepackage{amsmath,graphicx}
\usepackage[colorlinks=true,linktocpage=true,linkcolor=blue,citecolor=blue]{hyperref}
\usepackage[usenames,dvipsnames]{color}
\usepackage{float}
\usepackage{comment}
\usepackage{textcomp}
\usepackage[normalem]{ulem}
\usepackage{bm}

\begin{document}

\begin{frontmatter}



\dochead{XXVIIth International Conference on Ultrarelativistic Nucleus-Nucleus Collisions\\ (Quark Matter 2018)}

\title{Analytical solutions of causal relativistic hydrodynamic equations for Bjorken and Gubser flows}


\author[a]{Chandrodoy~Chattopadhyay  \footnote{Presenter. E-mail: chandrodoy.chattopadhyay@tifr.res.in}}
\author[b]{Amaresh~Jaiswal \tnoteref{tn0}}
\tnotetext[tn1]{ A.J. is supported in part by the DST-INSPIRE faculty award under Grant No. DST/INSPIRE/04/2017/000038.}
\author[a]{Sunil~Jaiswal}
\author[a]{Subrata~Pal}
\address[a]{Department of Nuclear and Atomic Physics, Tata Institute of Fundamental Research, Homi Bhabha Road, Mumbai 400005, India}
\address[b]{School of Physical Sciences, National Institute of Science Education and Research, HBNI, Jatni-752050, Odisha, India}

\begin{abstract}
We obtain general analytical solutions of third-order viscous hydrodynamic equations for Bjorken and Gubser flows in systems with vanishing bulk viscosity and chemical potential, and having a constant shear relaxation time. We also analytically characterize and determine the hydrodynamic attractors for such systems by studying the universal behavior of these solutions at both early and late times. Finally we study the properties of these hydrodynamic attractors for transport coefficients obtained from relativistic kinetic theory in the relaxation-time approximation.
\end{abstract}


\end{frontmatter}


\section{Introduction}
\label{}

The surprising success of relativistic viscous hydrodynamics in explaining flow data even in small systems (which are expected to be far from equilibrium) \cite{Bozek:2016jhf} formed in proton-lead (p-Pb) collisions or proton-proton (p-p) collisions has generated renewed interests in quantifying the domain of validity of hydrodynamics. Viscous hydrodynamics was originally formulated as an order-by-order expansion in gradients of the fluid four-velocity, under the assumption that these gradients normalised to the system's temperature are small.
While the first-order Navier-Stokes theory relates the shear tensor to instantaneous space-like velocity gradients \citep{Landau}, higher-order theories treat the dissipative quantities as an independent degree of freedom \cite{Israel:1979wp, Jaiswal:2013vta}. This leads to good agreement with the exact solution of the Boltzmann equation even in situations where the deviations from thermal equilibrium are large \cite{Jaiswal:2013vta, Jaiswal:2014isa}.
One can then explore whether these out-of-equilibrium situations can also be described merely by a gradient expansion.

Analytical solutions of higher-order dissipative hydrodynamics exist in very special cases \cite{Hatta:2014gqa, Denicol:2017lxn}. It is thus interesting to obtain analytical solutions and study the properties of these solutions such as dependence on initial conditions and late time behaviour. In this article, we consider two qualitatively different flow profiles, namely, Bjorken \cite{Bjorken:1982qr} and Gubser flows \cite{Gubser:2010ze}, in which the conservation equations are considerably simplified due to the symmetries inherent in these systems. We analytically solve the third-order hydrodynamic equation in Bjorken and Gubser flow and study the attractor behaviour of the solutions. For these analytical solutions, we demonstrate a method to prove the existence of the attractor solution by studying the late time behaviour and its characterization using the early time dynamics of the family of solutions. For both these flow profiles, we explore domains where gradients are large and study if the results for third-order hydrodynamics can also be described by perturbative expansions in terms of velocity gradients.


\section{Analytical solution for Bjorken flow} 
\label{ABS}

The third-order evolution equation for shear stress tensor in the case of one-dimensional Bjorken flow takes the form
\begin{align}
\label{BED} 
  \frac{d\epsilon}{d\tau} = -\frac{1}{\tau}\left(\frac{4}{3}\epsilon -\pi\right), \qquad
   \frac{d\pi}{d\tau} = - \frac{\pi}{\tau_\pi} 
   + \frac{1}{\tau}\left[\frac{4}{3}\beta_\pi - \left( \lambda + \frac{4}{3} \right) \pi - \chi\frac{\pi^2}{\beta_\pi}\right], 
\end{align}
where $\beta_\pi = 4P/5$, $\lambda = 10/21$ and $\chi = 72/245$. After some variable transformations, for constant shear relaxation time $\tau_\pi$, the above equations can be written as second-order linear ODE for $\bar{\pi}\equiv\pi/(\epsilon+P)$ as a function of $\bar{\tau}\equiv\tau/\tau_\pi$, which admits a solution in terms of Whittaker functions $M_{k,m}(\bar{\tau})$ and $W_{k,m}(\bar{\tau})$ \cite{Denicol:2017lxn, Jaiswal:Sunil},
\begin{equation}\label{CHI}
\bar{\pi}(\bar{\tau})= \frac{(2 k+2 m+1) M_{k+1,m}(\bar{\tau} )-2 \alpha W_{k+1,m}(\bar{\tau} )}{2 \gamma \ \left[ M_{k,m}(\bar{\tau} )+\alpha W_{k,m}(\bar{\tau }) \right]}, ~~~~ 
\epsilon(\bar{\tau})=
\epsilon_0 \left(\frac{\bar{\tau}_{0}}{\bar{\tau}}\right)^{\!\frac{4}{3} \left(1-\frac{k}{\gamma}\right)}\! \left[e^{-\frac{\left(\bar{\tau}-\bar{\tau}_{0}\right)}{2}} \frac{M_{k,m}(\bar{\tau} ) + \alpha W_{k,m}(\bar{\tau} ) } {M_{k,m}(\bar{\tau}_{0} ) + \alpha W_{k,m}(\bar{\tau}_{0} )} \right]^{\frac{4}{3 \gamma}}\!\!\!\!,
\end{equation} 
where $k\equiv-\frac{\lambda+1}{2},~ m\equiv\frac{1}{2}\sqrt{4a \gamma+\lambda^2}$ with $a\equiv\beta_\pi/\epsilon=4/15$ and $\gamma\equiv 4/3+5\chi=412/147$. In the above equation, $\epsilon_0$ is the energy density at initial time $\bar{\tau}_0$ and $\alpha$ determines the initial value for $\bar{\pi}$.


\subsection{Emergent attractor behaviour in Bjorken flow}\label{attractor}

We will now investigate the late time behaviour of the solution of $\bar{\pi}(\bar{\tau})$ obtained in the previous section. The motivation of this analysis is to check the existence of an attractor on which $\bar{\pi}$ with different initial conditions (encoded in $\alpha$) approaches at late times. The rate of convergence (if at all) of curves with different initial conditions can be obtained by taking a partial derivative with respect to $\alpha$. At late times we find that the separation between curves damp exponentially: 
$\partial \bar{\pi}/\partial \alpha \sim e^{-\bar{\tau}}/\bar{\tau}$
in the limit  $\bar{\tau} \to \infty$. This shows that the separation between solutions corresponding to different initial conditions damps exponentially with a time-scale which is equal to the shear relaxation time.

Since we have now confirmed the existence of attractor for  $\bar{\pi}(\bar{\tau})$, we will go ahead and look for it's solution.
From Eq.~(\ref{CHI}), it is easy to see that the solution of $\bar{\pi}(\bar{\tau})$ completely loses information about initial conditions (encoded in $\alpha$) as Whittaker $M_{k,m}(\bar{\tau})$ dominates Whittaker $W_{k,m}(\bar{\tau})$ at late times. So the solution which exactly reproduces the late time behavior of Eq.~(\ref{CHI}) has $\alpha = 0$, and is {\em defined} to be the hydrodynamic attractor \cite{Denicol:2017lxn}. In our case,
this yields 
\begin{equation}\label{CHI1}
\bar{\pi}_{att}(\bar{\tau})= \left(\frac{2k + 2m + 1 }{2 \gamma }\right) \  \frac{M_{k+1,m}(\bar{\tau} )}{M_{k,m}(\bar{\tau} )}.
\end{equation}
We propose a different method of characterizing the attractor solution which is based on the idea that at early times $\bar{\tau} \sim 0$ the attractor is the unique solution which connects to the 
positive branch, whereas, all other solutions converge to a negative value (see Eq. (\ref{BED}) at $\bar{\tau} \sim 0$). This suggests that the quantity 
$\psi(\alpha) \equiv \lim_{\bar{\tau} \to 0} \partial \bar{\pi}/ \partial \alpha$ should possess a singularity for some value of $\alpha$ which corresponds to the attractor. Using the expression for $\bar{\pi}(\alpha,\bar{\tau})$
from Eq.~(\ref{CHI}), we find that $\psi(\alpha)$ indeed diverges at $\alpha = 0$, thus proving that Eq.~(\ref{CHI1}) is the hydrodynamic attractor for third-order theory.

\begin{figure}[t]
 \begin{center}
  \scalebox{.4}{\includegraphics{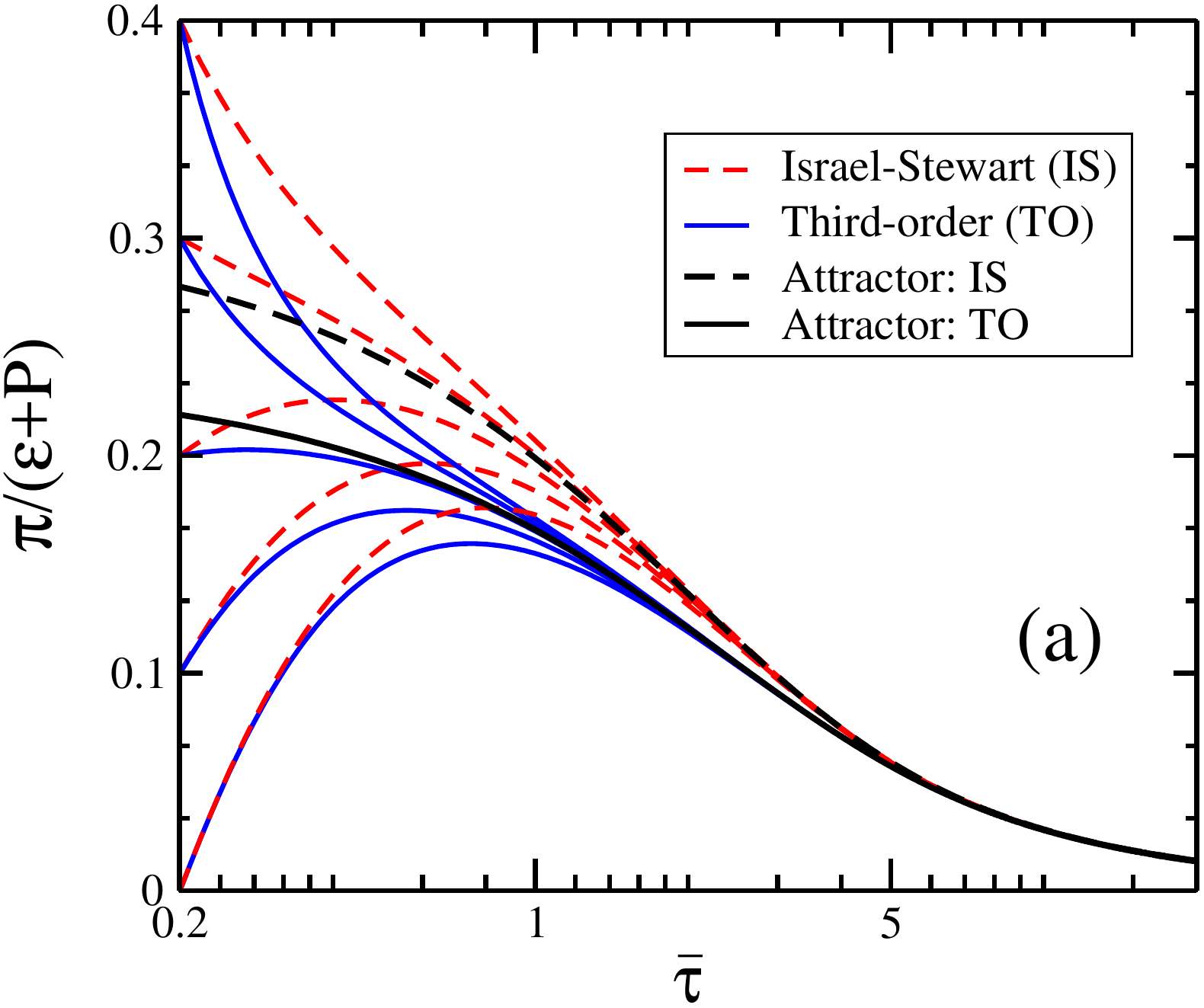}}\hfil
  \scalebox{.4}{\includegraphics{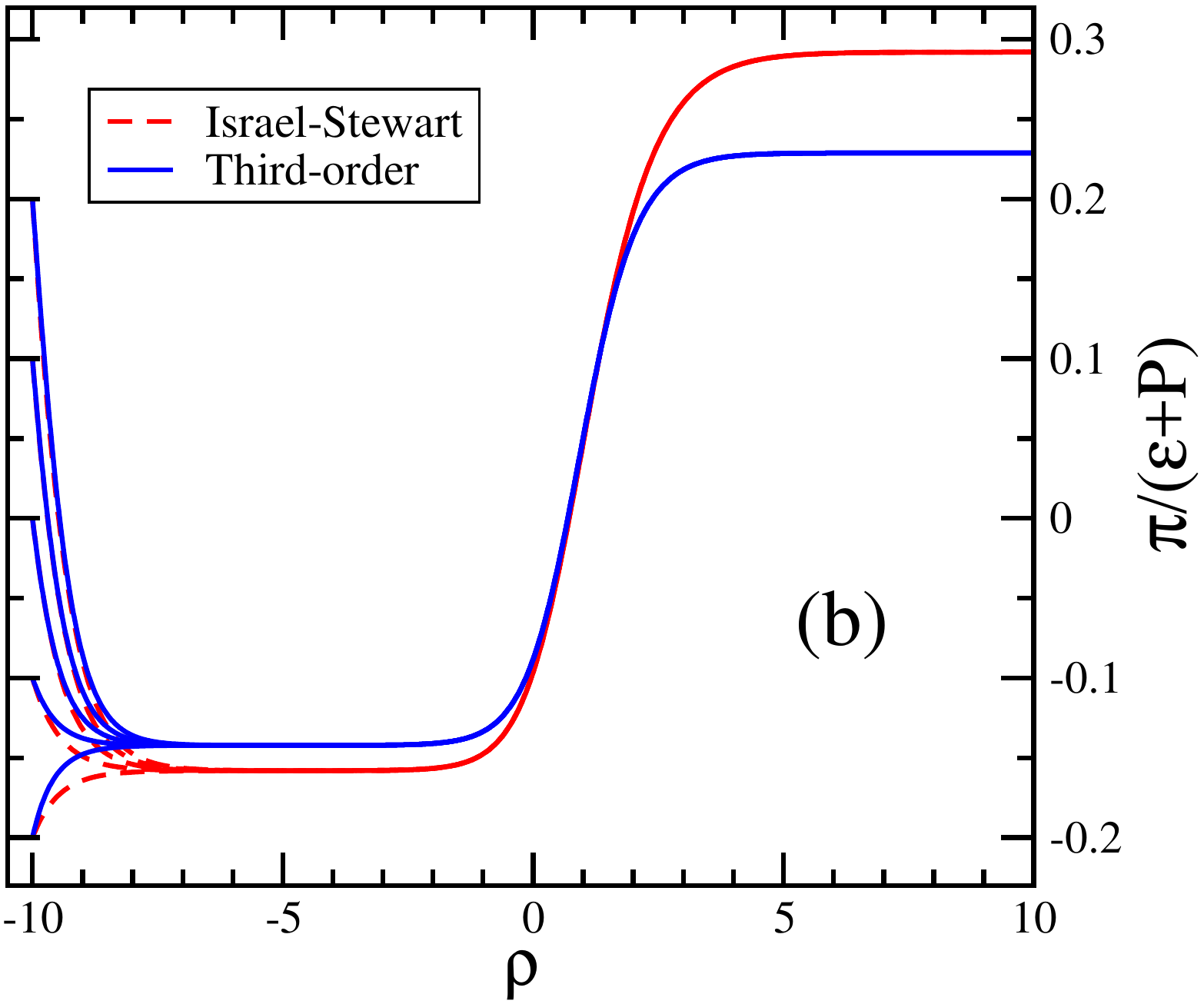}}
 \end{center}
 \vspace{-0.8cm}
 \caption{(a): Variation of $\bar{\pi}$ with $\bar{\tau}$ of Eq.~(\ref{CHI}), plotted for different initial conditions (initial $\bar{\tau}_{0}=0.1$), converges on the attractor (Eq.~(\ref{CHI1})) before local equilibrium is reached. (b): Variation of $\hat{\bar{\pi}}$ with $\rho$ of Eq.~(\ref{GCHI}), plotted for different initial conditions for Israel-Stewart and third-order theories.}
 \label{NSHEAR}
\end{figure}

In Fig.~\ref{NSHEAR} (a), we show how normalised shear with different initial conditions evolve in $\bar{\tau}$ for second and third order hydrodynamic theory. Dashed red curves are solutions to second-order Israel-Stewart theory and solid blue curves are those for third-order hydrodynamics obtained from Chapman-Enskog like expansion. We see that there are different attractors for the Israel-Stewart and third-order hydrodynamics represented by black dashed and solid curves, respectively. At late times ($\bar{\tau}\gtrsim 5$), we see that the solutions corresponding to Israel-Stewart and third-order theories merge indicating convergence to Navier-Stokes solution.

\section{Analytical solution for Gubser flow}

In this section, we present for the first time analytic solution of Gubser flow for third order assuming constant shear relaxation time $\hat{\tau}_{\pi}$ approximation. The evolution equations for the Weyl rescaled variables
\cite{Gubser:2010ze} (denoted by `hats'), $\hat{\epsilon}$ and $\hat{\pi}$, in de Sitter coordinates take the form \cite{Chattopadhyay:2018apf}
\begin{align}
\label{GED}
   \frac{d\hat{\epsilon}}{d\rho} = - \left( \frac{8}{3} \hat{\epsilon}  - \hat{\pi} \right) \tanh\rho, \qquad
   \frac{d\hat{\pi}}{d\rho} = - \frac{\hat{\pi}}{\hat{\tau}_\pi} 
   + \tanh\rho\left(\frac{4}{3} \hat{\beta}_\pi - \hat{\lambda} \hat{\pi} - \hat{\chi}\frac{\hat{\pi}^2}{\hat{\beta}_\pi}\right),
\end{align} 
where $\hat{\beta}_\pi = 4\hat{P}/5$, $\hat{\lambda} = 46/21$ and $\hat{\chi} = 72/245$. After some variable transformations, the above equations can be written as second-order linear ODE for $\hat{\bar{\pi}}\equiv \hat{\pi}/(\hat{\epsilon} + \hat{P})$ as a function of $\rho$ which has four regular singular points at $(-1,0,1,\infty)$ \citep{Jaiswal:Sunil}. The general solution of this linear ODE are in terms of Heun general function (H) and its derivative (H'):
\begin{align}
 \nonumber
	 \hat{\bar{\pi}}(\rho) =&~ \bigg[ \hat{\alpha}~ {\mathrm{T}_{+}}^{-(a_{1}/4)}  \bigg( \Big( 1+(\lambda + {a_1}){\hat{\tau}}_\pi \Big) ~\mathrm{\mathrm{T}}_{-}  - 4 {\hat{\tau}}_\pi b_4 \mathrm{T}_{+} \bigg) ~\mathrm{H}(s) 	 
	 + {\mathrm{T}_{+}}^{(a_{1}/4)}  \bigg( \Big( 1+(\lambda - {a_1}){\hat{\tau}}_\pi \Big) ~\mathrm{T}_{-}  - 4 {\hat{\tau}}_\pi {b}_4 \mathrm{T}_{+} \bigg) ~\mathrm{H}(r) 
	 \\ 
	 &- 4 {\hat{\tau}}_\pi \mathrm{T}_{-}\bigg( \hat{\alpha}~ {\mathrm{T}_{+}}^{1-(a_{1}/4) } ~\mathrm{H}'(s) + {\mathrm{T}_{+}}^{1+(a_{1}/4)} ~\mathrm{H}'(r) \bigg) \bigg] ~\bigg/~
	  \left[ 4 \gamma \tanh\rho \hat{\tau}_\pi \bigg( \hat{\alpha}~ {\mathrm{T}_{+}}^{-(a_{1}/4)} \mathrm{H}(s) + {\mathrm{T}_{+}}^{(a_{1}/4)} \mathrm{H}(r) \bigg) \right], 	  
\label{GCHI}\\
\hat{\epsilon}(\rho)=&~ \hat{\epsilon}_0 \left(\frac{\cosh\rho_0}{\cosh\rho}\right)^{8/3} \left(\frac{T_{-}}{T_{0-}}\right)^{\frac{4 b_4}{3 \gamma}} \left(\frac{\mathrm{T}_{+}}{T_{0+}} \right)^{-\frac{1}{3 \gamma} \left( \frac{1}{{\hat{\tau}}_{\pi}} + \lambda \right)} 
 ~ \left(\frac{\hat{\alpha} \mathrm{T}_{+}^{-(a_1 /4)} ~\mathrm{H}(s) + \mathrm{T}_{+}^{(a_1 /4)} ~\mathrm{H}(r_0)}{\hat{\alpha} T_{0+}^{-(a_1 /4)} ~\mathrm{H}(s_0) + T_{0+}^{(a_1 /4)} ~\mathrm{H}(r_0)}\right)^{\frac{4}{3 \gamma}}, \label{GENGY}
\end{align}
where,
\begin{align*}\label{HEUN}
r\equiv& \left\lbrace 2,\frac{b_3 - b_2}{24 ~a_3},\frac{a_1 + a_3}{4},\frac{a_1 a_3 b_1}{4~ a_3},\frac{a_1}{2}+1,-1,T_{+} \right\rbrace,\quad 
s \equiv \left\lbrace 2,\frac{b_3 + b_2}{24 ~a_3},\frac{-a_1 + a_3}{4},\frac{-a_1 a_3  b_1}{4~ a_3},\frac{a_1}{2}+1,-1,T_{+} \right\rbrace,
\end{align*}
denotes the seven arguments required for H functions with $T_{\pm} \equiv \tanh\rho \pm 1$. $r_0$ and $s_0$ is defined as above at $\rho=\rho_0$ with $T_{0\pm} \equiv \tanh\rho_{0} \pm 1$, and $\hat{\epsilon}_0$ is the energy density at $\rho_0$.
Here $a_1, a_2, a_3, b_1, b_2, b_3$ and $b_4$ are constants that depends on the transport coefficients and $\hat{\tau}_{\pi}$:
\begin{align*}
a_1 & \equiv \frac{1}{\hat{\tau}_{\pi}} \sqrt{1 + \left(4 a \gamma + {\lambda}^{2} \right) {\hat{\tau}_{\pi}}^2 + 2 \lambda \hat{\tau}_{\pi}}, \qquad\qquad\quad
a_2 \equiv \frac{1}{\hat{\tau}_{\pi}} \sqrt{\left(4 a \gamma + {\lambda}^{2} \right)\left(\left(4 a \gamma + {\lambda}^{2} \right){\hat{\tau}_{\pi}}^2 -2 \lambda \hat{\tau}_{\pi} + 1 \right)},&
\nonumber \\ 
a_3 & \equiv \frac{1}{\hat{\tau}_{\pi}} \sqrt{1 + \left(20 a \gamma + 5{\lambda}^{2} + 4 a_2 \right) {\hat{\tau}_{\pi}}^2 - 2 \lambda \hat{\tau}_{\pi}}, \quad~~
b_1 \equiv \frac{1}{{\hat{\tau}_{\pi}}^2} \left(-3\left(4 a \gamma + {\lambda}^{2} \right){\hat{\tau}_{\pi}}^2 -2 \lambda \hat{\tau}_{\pi} + 1 \right),&
\nonumber \\ 
b_2 & \equiv \frac{-a_1}{{\hat{\tau}_{\pi}}^2} \left(4+ \left(6 a_2 - 6 a_3 - b_1 \right){\hat{\tau}_{\pi}}^2 -8 \lambda \hat{\tau}_{\pi} \right), \quad\quad
b_4 \equiv \frac{1}{12 a_3 {\hat{\tau}_{\pi}}^2} \left( 4 + \left(-3 a_3 \lambda - b_1 + 6 a_2 \right){\hat{\tau}_{\pi}}^2 + \left(3 a_3 - 8 \lambda \right)\hat{\tau}_{\pi} \right), \nonumber \\ 
b_3 & \equiv \frac{1}{{\hat{\tau}_{\pi}}^2} \left( \left( a_3 b_1 + 12 a_2 - 2 b_1 \right){\hat{\tau}_{\pi}}^2 + \left( 2 \left( \lambda + 6\right) a_3 -16 \lambda \right) \hat{\tau}_{\pi} + 2 a_3 + 8 \right). 
\end{align*}
where $a\equiv\hat{\beta}_\pi/\hat{\epsilon}=4/15$, $\lambda\equiv 8/3-\hat{\lambda}=10/21$ and $\gamma\equiv 4/3+5 \hat{\chi}=412/147$.

From Eqs.~(\ref{GCHI}) and (\ref{GENGY}), we see that $\hat{\bar{\pi}}(\rho)$ and $\hat{\alpha}$ can be found for initial values from Eq.~(\ref{GCHI}) and the relation $\hat{\bar{\pi}} \equiv \hat{\pi}/(\hat{\epsilon} + \hat{P})$. So this initial value problem is defined by the constants $\hat{\epsilon}_0$ and $\hat{\alpha}$. The above results hold for second-order Israel-Stewart theory as well because form of the differential equation remains the same. The only change is the coefficient $\gamma$ taking the value $4/3$. In Fig.~\ref{NSHEAR} (b), we show the solutions of normalised shear for various initial conditions for Israel-Stewart theory represented by red curves and third-order theory in blue curves. We see that the solutions for these two theories converge separately indicating the existence of attractor solutions. The characterization of the attractor solutions for Gubser flow profile using the definition introduced in Sec.~\ref{attractor} is left for future work \cite{Jaiswal:Sunil}.


\section{Gradient expansion as a perturbation series}

We now investigate both flow profiles in domains where gradients are large and study if the results for third-order hydrodynamics can also be described by series expansions in terms of velocity gradients.

For Bjorken flow, each term in the shear evolution equation are labelled according to the number of gradients they contain, with the normalised shear $\bar{\pi}$ and the gradient term $\theta = 1/\tau$ considered to be of order $\epsilon$ and we look for a series solution in power of $\epsilon$ for $\bar{\pi}$ , 
\begin{equation}\label{GEB2} 
\epsilon \frac{d\bar{\pi}}{d\bar{\tau}} + \bar{\pi} + \epsilon \frac{\lambda\bar{\pi}}{\bar{\tau}} - \frac{a}{\bar{\tau}} 
	+ \epsilon^2 	\frac{\gamma \bar{\pi}^2}{\bar{\tau}} = 0, \qquad
	 \bar{\pi} = \sum_{n=0}^{\infty} \bar{\pi}_{n} (\bar{\tau}) \epsilon^{n}.
\end{equation}
The solution of these infinite equations are given by the recurrence relation
\begin{equation}
\bar{\pi}_{0} = \frac{a}{\bar{\tau}}, \qquad
\underset{n \geq 1}{\bar{\pi}_{n}} = - \left[ \frac{d \bar{\pi}_{n-1}}{d \bar{\tau}} + \frac{\lambda}{\bar{\tau}} \bar{\pi}_{n-1} + \frac{\gamma}{\bar{\tau}} \sum_{m=0}^{n-2}   \bar{\pi}_{n-m-2} \ \bar{\pi}_{m}   \right].
\end{equation}
Using the same method to label the terms of normalized shear equation for Gubser flow we have,
%
\begin{align}
\epsilon \frac{d\hat{\bar{\pi}}}{d\rho} = - \frac{\hat{\bar{\pi}}}{\hat{\tau}_\pi} 
	  + \tanh\rho \left[ a + \epsilon \lambda \hat{\bar{\pi}} - {\epsilon}^2 \gamma  \hat{\bar{\pi}}^2  \right], \qquad
	  \hat{\bar{\pi}} = \sum_{n=0}^{\infty} \hat{\bar{\pi}}_{n} (\hat{\tau}) \epsilon^{n}.	\label{GGE2} 
\end{align}
The solution for these equations are given by the relation
\begin{align}\label{GGEE1}
\hat{\bar{\pi}}_0 =&~ a \, \hat{\tau}_\pi \tanh \rho, \qquad
\hat{\bar{\pi}}_1 = a \, \hat{\tau}_\pi^2 \left[ \tanh^2 \rho ~(1+\lambda) - 1 \right], ~~\\
\underset{n \geq 2}{\hat{\bar{\pi}}_n} =&~ -\hat{\tau}_\pi \left[ \frac{d\hat{\bar{\pi}}_{n-1}}{d\rho} - \lambda \tanh(\rho) ~\hat{\bar{\pi}}_{n-1} + \gamma \tanh \rho \sum_{m=0}^{n-2} \hat{\bar{\pi}}_{n-m-2} \hat{\bar{\pi}}_m \right].
\end{align}
The gradient series was found to diverge in both Bjorken and Gubser flows for larger values of $n$.
Even the slow roll expansion obtained by assuming slow variation of $\bar{\pi}$ with 
`time' ($\tau$ in Bjorken and $\rho$ in Gubser) shows large oscillations, suggesting the failure of these traditional perturbative schemes in properly characterizing 
the relative contributions of various terms in the shear evolution equation.










\end{document}